%% file: n613.tex
\documentclass[twocolumn]{aastex631}
\usepackage{ulem}

\input{include.tex}
\hypersetup{linkcolor=red,citecolor=green,filecolor=cyan,urlcolor=magenta}

\shorttitle{Distributions of the Density and Temperature of the Molecular Gas of NGC\,613}
\shortauthors{Kaneko et al.}

\begin{document}
	
	\title{Distributions of the Density and Kinetic Temperature of the Molecular Gas \\in the Central Region of NGC\,613 using Hierarchical Bayesian Inference}
	
	\correspondingauthor{Hiroyuki Kaneko}
	\email{kaneko.hiroyuki.astro@gmail.com}

	\author[0000-0002-2699-4862]{Hiroyuki Kaneko}
	\affiliation{Joetsu University of Education, 1, Yamayashiki-machi, Joetsu, Niigata 943--8512, Japan}
	\affiliation{National Astronomical Observatory of Japan, 2-21-1 Osawa, Mitaka, Tokyo 181--8588, Japan}
	\affiliation{Center for Astronomy, Ibaraki University, 2-1-1 Bunkyo, Mito, Ibaraki 310--8512, Japan}
	\affiliation{Institute of Science and Technology, Niigata University, 8050, Ikarashi 2-no-cho, Nishi-ku, Niigata, 950-2181 Japan}
	
	\author[0000-0001-9016-2641]{Tomoka Tosaki}
	\affiliation{Joetsu University of Education, 1, Yamayashiki-machi, Joetsu, Niigata 943--8512, Japan}
	
	\author[0000-0001-8153-1986]{Kunihiko Tanaka}
	\affiliation{Department of Physics, Faculty of Science and Technology, Keio University, 3-14-1, Hiyoshi, Yokohama, Kanagawa 223--8522, Japan}
	
	\author[0000-0002-7616-7427]{Yusuke Miyamoto}
	\affiliation{Department of Electrical and Electronic Engineering, Faculty of Engineering, Fukui University of Technology, 3-6-1, Gakuen, Fukui, Fukui, 910--8505, Japan}

	\begin{abstract}
		We present position-position-velocity (PPV) cubes of the physical and chemical properties of the molecular medium in the central 1.2 kpc region of the active galaxy \object{NGC\,613} at a PPV resolution of 0$\farcs$8 $\times 0\farcs8\times10\ \mathrm{km\,s^{-1}}$ (0$\farcs$8 = $\sim$68\,pc).
		We used eight molecular lines ($^{13}$CO(1-0), C$^{18}$O(1-0), HCN(1-0), HCO$^{+}$(1-0), CS(2-1), HCN(4-3), HCO$^{+}$(4-3), and CS(7-6)) obtained with ALMA. 
		Non-LTE calculation with hierarchical Bayesian inference was used to construct PPV cubes of the gas kinetic temperature ($T_\mathrm{kin}$), molecular hydrogen volume density ($n_\mathrm{H_2}$), column densities ($N_\mathrm{H_2}$), and fractional abundances of four molecules ($^{12}$C$^{18}$O, HCN, HCO$^{+}$, and CS).
		The derived $n_\mathrm{H_2}$, $N_\mathrm{H_2}$, and $T_\mathrm{kin}$ ranged 10$^{3.21-3.85}$ cm$^{-3}$, 10$^{20.8-22.1}$ cm$^{-2}$, and 10$^{2.33-2.64}$ K, respectively. 
		Our first application of the non-LTE method with the hierarchical Bayesian inference to external galaxies yielded compatible results compared with the previous studies of this galaxy, demonstrating the efficacy of this method for application to other galaxies.
		We examined the correlation between gas surface density $\Sigma_\mathrm{H_2}$ (converted from $N_\mathrm{H_2}$) and the star formation rate $\Sigma_\mathrm{SFR}$ obtained from the 110 GHz continuum flux map and found two distinct sequences in the $\Sigma_\mathrm{H_2}$-$\Sigma_\mathrm{SFR}$ diagram; 
		the southwestern subregion of the star-forming ring exhibited a $\sim$0.5 dex higher star formation efficiency (SFE; $\Sigma_\mathrm{SFR}$/$\Sigma_\mathrm{H_2}$) than the eastern subregion. 
		However, they exhibited no systematic difference in $n_\mathrm{H_2}$, which is often argued as a driver of SFE variation. 
		We suggest that the deficiency of molecular gas in the southwestern subregion, where no significant gas supply is evident along the offset ridges in the bar, is responsible for the elevated SFE. 
	\end{abstract}
	

\section{Introduction} \label{sec:intro}
Giant molecular clouds (GMCs) with a typical size of tens of parsecs \citep{Scoville1987} have been regarded as the formation sites of massive stars, which are a major driver of the physical and chemical evolution of galaxies. 
It is essential to investigate what kind of physical processes govern the behavior of GMCs, and thus star formation in galaxies. 
This can be addressed by measuring the physical properties of the molecular gas, including the gas kinetic temperature ($T_\mathrm{kin}$), molecular hydrogen gas volume, and surface densities ($n_\mathrm{H_2}$ and $\Sigma_\mathrm{H_2}$, respectively) in galaxies. 
In particular, cloud-averaged $n_\mathrm{H_2}$, or a dense-gas fraction, is crucial to studying star-formation law in galaxies. 
HCN and HCO$^{+}$ emission lines are used as a dense ($n_\mathrm{H_2} \gtrsim 10^4$ cm$^{-3}$) molecular gas tracer, despite some caveats from studies of Galactic clouds \citep[see, e.g.,][]{Pety2017, Nishimura2017}.
Star formation rates (SFRs) are traced by, for example, far-infrared (FIR) luminosities ($L_\mathrm{FIR}$).
A strong correlation between dense ($n_\mathrm{H_2} \gtrsim 10^4$ cm$^{-3}$) molecular gas and SFRs is observed on a galaxy scale in star-forming galaxies and luminous infrared galaxies in the local universe \citep{GaoSolomon2004,GarciaBurillo2012}.
Furthermore, the relation between the dense-gas fraction and star formation may depend on a $\gtrsim 1$ kpc-scale environment \citep{Usero2015,Bigiel2016}. 
The next step is to measure the physical properties of galaxies with a GMC-scale resolution and relate them to a large-scale environment. 

Previous (and even current) studies of GMCs in galaxies have relied on single or few-line measurements to identify the structure and the physical properties of the GMCs \citep[e.g.,][]{Hughes2013}.
Many physical and chemical quantities have been investigated with those measurements, such as the density, temperature, pressure, ionization degree, and molecular abundances of various species \citep[e.g.,][]{Tosaki2017, Herrera2020}.
For example, PHANGS-ALMA \citep{Rosolowsky2021} uses $^{12}$CO(2--1) emission to trace the distribution and mass of molecular gas.
However, the range of information accessible through single or few-line emissions reflects just one aspect of GMCs, when molecular gas is clearly physically and chemically complex. 
Now, with the advent of the Atacama Large Millimeter/submillimeter Array (ALMA), it has become possible to obtain data on multiline emissions in galaxies with similar spatial resolution and sensitivity to previous observations made in the star-forming regions in the Milky Way.

To understand how the properties of GMCs affect the star formation activity in a galaxy, we focus on GMCs in star-forming rings around active galactic nuclei (AGN).
Two scenarios explain star formation in a star-forming ring \citep{Boker2008}.
The first one is called the ``popcorn-like'' star formation.
In this scenario, star formation takes place stochastically within the ring or occurs in the entire ring at the same time.
This confers no systematic age sequence of formed stars.
The other is the ``pearls-on-a-string'' star formation.
According to the ``pearls-on-a-string'' scenario, overdensity gas clouds move with the rotation of the ring, and continuously form star clusters along the ring.
Star formation is ceased due to a first supernova or gas consumption.
Thus, a sequential age difference of star clusters is expected in this scenario.
Optical and infrared observations reported that both sequential and nonsequential age gradient cases exist.
For example, \citet{Mazzuca2008} showed a significant age gradient in 10 out of 21 star-forming rings.
On the other hand, the star-forming ring of NGC\,1672 does not show an age gradient, indicating popcorn-like star formation \citep{Fazeli2020}. 
\citet{Rico-Villas2021} observed NGC\,1068 with multiple HC$_{3}$N lines and concluded that star formation of the star-forming ring in NGC\,1068 could be explained by the ``popcorn'' scenario.
However, these observations have not always included the physical properties of GMCs, which serve as the direct ingredients in star formation.
Our goal herein is to understand the starburst-ring environment by measuring the physical properties of GMCs.

For this purpose, we focus on the central region of the nearby barred galaxy \object{NGC\,613}.
According to \citet{Boker2008}, this galaxy hosts a low-luminosity AGN and a star-forming ring.
The central nuclear disk (CND) has a diameter of $\sim$100\,pc, while the star-forming ring has a diameter of $\sim$600\,pc.
The Hubble Space Telescope WFPC2 image (PI: Stephan Smartt) shows that dust connects the bar structure and the star-forming rings.
Such a distribution is also observed in molecular gas both in low-density gas tracers (e.g. $^{13}$CO~$J$=1--0) and in high-density tracers (e.g. CO~$J$=3--2 and HCO$^{+}$~$J$=1--0; \citealt{Miyamoto2017}).
Their velocity structure indicates that molecular gas is supplied to the star-forming ring along the bar.
Active star formation in the central region of NGC\,613 is only seen in the star-forming ring, which is traced by Br$\gamma$ \citep{Boker2008} and free-free radio continuum \citep{Miyamoto2017}, and NGC\,613 experiences pearls-on-a-string-like star formation in its star-forming ring \citep{Falcon-Barroso2014}.
Considering the proximity to the galaxy \citep[17.5~Mpc;][]{Tully1988} and the moderate inclination angle of the star-forming ring \citep[55$^\circ$;][]{HummelJorsater1992}, these properties indicate that the central region of NGC\,613 is an ideal candidate for our study.

In this study, we present position--position--velocity (PPV) 3D distributions (hereafter ``3D distributions'') of gas properties, including molecular hydrogen volume density $n_\mathrm{H_2}$, molecular hydrogen column density $\Sigma_\mathrm{H_2}$, and gas kinematic temperature $T_\mathrm{kin}$ in the central 1.2 kpc region at spatial and spectral resolutions of 0$\farcs$8 ($\sim$68~pc) and 10~km~s$^{-1}$, respectively.
We exploit rich multimolecular observations using ALMA, including $^{13}$CO($J$=1--0), C$^{18}$O($J$=1--0), HCN($J$=4--3), HCN($J$=1--0), HCO$^{+}$($J$=4--3), HCO$^{+}$($J$=1--0), CS($J$=7--6), and CS($J$=2--1) obtained by \citet{Miyamoto2017,Miyamoto2018}.
\citet{Audibert2019} also observed CO($J$=3--2), HCN($J$=4--3), HCO$^{+}$($J$=4--3), and CS($J$=7--6) emission lines toward the same region with a higher resolution ($\sim 0\farcs2$).
A similar combination of higher-density tracers (HCN, HCO$^{+}$, and CS) and less-dense gas tracers (CO and its isotopologueues) have been obtained in the central regions of local galaxies using ALMA and NOEMA \citep[e.g.,][]{Viti2014,Saito2015,Salak2018,Beslic2021}.
These data have been utilized to constrain $n_\mathrm{H_2}$ and $T_\mathrm{kin}$ via non-local thermodynamic equilibrium (non-LTE) analysis using, for example, RADEX \citep{vanderTak2007}. 
However, previous studies used a limited number of low-$J$ transitions, which resulted in a biased measurement of $n_\mathrm{H_2}$. 
To prevent this bias, we derive physical quantities from these multiple-molecular-line observations by exploiting a non-LTE method with hierarchical Bayesian (HB) inference.
This method has been successfully demonstrated using many molecular lines in the central molecular zone in the Milky Way \citep[][hereafter T18]{Tanaka2018}.

The remainder of this paper is structured as follows. 
First, we describe the ALMA data and methodology used to obtain physical quantities via a non-LTE method with HB inference in section \ref{sec:data}. 
Second, the derived 3D views of $n_{\rm H_2}$, $\Sigma_{\rm H_2}$ and $T_{\rm kin}$ in NGC\,613 are presented in section \ref{sec:results}, and their relation to star formation properties is discussed in section \ref{sec:discuss}.
Finally, we summarize the outcomes of this study in section \ref{sec:summary}.

\section{ALMA data and analysis}\label{sec:data}

\subsection{Molecular Gas Data}
\begin{figure*}[htp]
	\begin{center}
		\includegraphics[width=160mm]{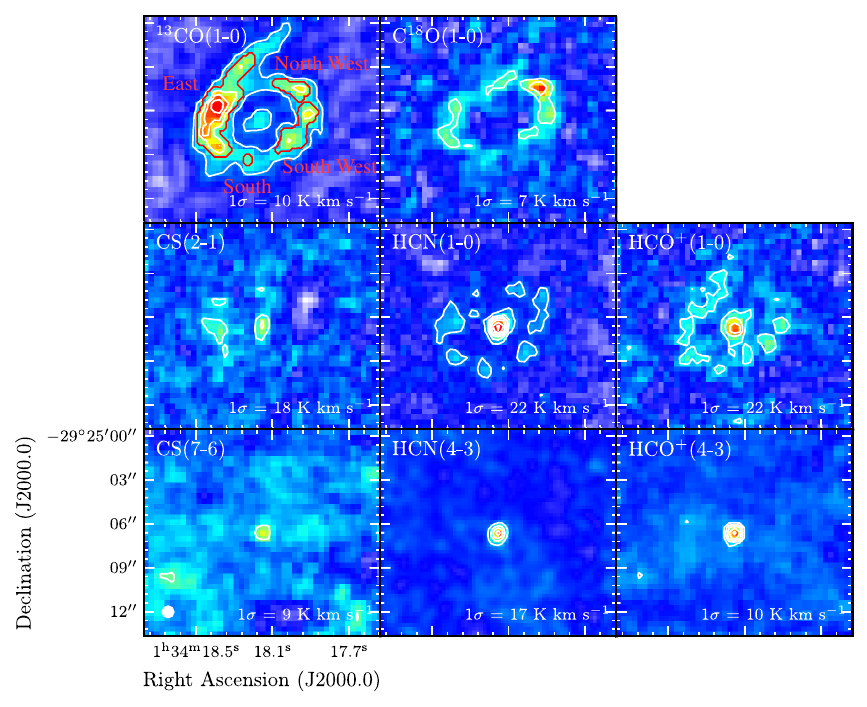}
	\end{center}
	\caption{Integrated intensity maps of eight molecular lines. The contour interval is 3$\sigma$, and	1$\sigma$ for each line is indicated at the bottom of the panels. All maps have an angular resolution of 0$\farcs$8 (beam size is indicated by a white circle in the bottom-left corner of CS(7--6)). Four subregions (east, south, southwest, and northwest) are defined based on the 6$\sigma$ contour of the $^{13}$CO integrated intensity, shown by the red contour in the top-left panel.}
	\label{fig:maps}
\end{figure*}

In this study, we used CO lines including isotopologueues ($^{13}$CO~($J$=1--0) and C$^{18}$O~($J$=1--0)) and multiple transitions of HCN, HCO$^{+}$, and CS~($J$=1--0 and 4--3 for HCN and HCO$^{+}$, and $J$=2--1 and 7--6 for CS).
All HCN, HCO$^{+}$, and CS data were acquired with ALMA by \citet{Miyamoto2017} (2013.1.01329.S).
We did not use the CO($J$=3--2) line observed by \citet{Miyamoto2017} because CO(3--2) is highly optically thick and close to thermalization, and hence violates the assumption of a common beam-filling factor for all input lines.
Breakdown of this assumption with low-$J$ CO lines critically affects the excitation analysis, as their absolute intensities immediately determine $T_\mathrm{kin}$.
Similarly, we did not use the datasets of \citet{Audibert2019} to avoid a biased measurement of gas properties in this study since they do not include low-density-tracer datasets such as $^{13}$CO($J$=1--0) and C$^{18}$O($J$=1--0) with a similar resolution.
The $^{13}$CO($J$=1--0) and C$^{18}$O($J$=1--0) data were obtained with ALMA by \citet{Miyamoto2018} (2015.1.01487.S).
The data details are described in \citet{Miyamoto2017,Miyamoto2018}.
All data were obtained with the 12 m array, the 7 m array, and the Total Power Array so that they do not suffer from missing flux problems.
Figure~\ref{fig:maps} illustrates the integrated intensity maps of the lines that we used.
In order to discuss the relationship between the GMC properties and SFR among the different parts of the ring, we divide the ring into four representative subregions (east, south, southwest, and northwest) based on the $^{13}\mathrm{CO}$ integrated intensity map (indicated as red contours in figure~\ref{fig:maps}).
The frequencies, upper-state energies ($E_\mathrm{u}$), rms noise, and field of view of the original observations of the target lines are listed in table~\ref{tab:spe_para}.

\begin{deluxetable*}{ccrrccr}
	\tablecaption{Spectral Parameters Used for This Study\label{tab:spe_para}}
	\tablehead{\colhead{Molecule} & \colhead{Transition} & \colhead{Frequency\tablenotemark{*}} & \colhead{$E_\mathrm{u}/k$\tablenotemark{$\dagger$}} & \colhead{rms \tablenotemark{$\ddagger$}} & \colhead{Field of View\tablenotemark{$\S$}} & \colhead{rms at $0\farcs8$ Resolution\tablenotemark{$\|$}}\\
		~ & ~ & (GHz) & (K) & (mJy beam$^{-1}$) & (arcsec) & (mK)
	}
	\decimals
	\startdata
	$^{13}$CO	&	$J$=1--0	&	110.201	&	5.29 	& 0.29 & 32 & 75\\
	C$^{18}$O	&	$J$=1--0	&	109.782	&	5.27    & 0.28 & 32	& 68\\
	HCN         &	$J$=1--0	&	88.632	&	4.25	& 0.73 & 62 & 220\\
	~           &	$J$=4--3	&	354.505	&	42.53	& 0.19 & 25 & 70\\
	HCO$^{+}$   &	$J$=1--0	&	89.189	&	4.28	& 0.75 & 62 & 215\\
	~			&	$J$=4--3	&	356.734	&	42.80	& 0.16 & 25 & 63\\
	CS			&	$J$=2--l	&	97.981	&	7.05	& 0.71 & 62 & 212\\
	~			&	$J$=7--6	&	342.883	&	65.83	& 0.12 & 25 & 78
	\enddata
	\tablenotetext{*}{Sourced from NIST Recommended Rest Frequencies for Observed Interstellar Molecular Microwave Transitions (F.~J. Lovas~et~al.; \url{http://physics.nist.gov/cgi-bin/micro/table5/start.pl})}
	\tablenotetext{\dagger}{The Cologne Database for Molecular Spectroscopy \citep[CDMS:][]{Muller2005}}
	\tablenotetext{\ddagger}{The values are derived from the original data \citep{Miyamoto2017,Miyamoto2018}.}
	\tablenotetext{\S}{Sourced from ALMA Science Archive (\url{https://almascience.nrao.edu})}
	\tablenotetext{\|}{The rms noise measured after convolving to $0\farcs8$. Details are provided in subsection \ref{subsection:Method:HBI}.  We note that the velocity width in deriving the rms noise is 10 km s$^{-1}$.}
\end{deluxetable*}

\subsection{Star Formation Data}
\label{SFR100GHz}
In previous studies, the near-infrared Br$\gamma$ emission was used as a tracer of star formation \citep[e.g., ][]{Miyamoto2017,Sato2021}.
Br$\gamma$ emission is a better tracer of dust-enshrouded star-forming regions than optical hydrogen recombination lines like H$\alpha$, but they are known to be limited by dust attenuation in the hearts of starburst galaxies \citep{Bendo2015, Bendo2016}.

\begin{figure}[htp]
	\begin{center}
		\includegraphics[width=8cm]{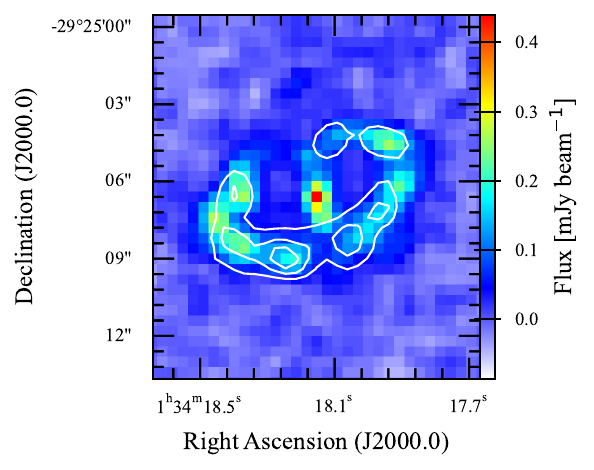} 
	\end{center}
	\caption{The 110 GHz continuum image of the central region of NGC\,613 with Br$\gamma$ contours.
		Both images were smoothed to a resolution of 0\farcs8 and regrided to a grid of 0\farcs4. 
		Contour levels are 10$\sigma$, 20$\sigma$, and 30$\sigma$, where 1$\sigma$ = 1.23 $\times$ 10$^{-15}$ erg s$^{-1}$ cm$^{-2}$ pixel$^{-1}$.}\label{fig:110GHz}
\end{figure}

This paper uses the 110 GHz continuum flux as an alternative star formation tracer since it is free from dust absorption.
\citet{Scoville1991} suggested that the 110 GHz continuum is related to the ionizing photon production rate, which is correlated to the O star luminosity, assuming that the 110 GHz continuum is entirely from free-free emission.
Based on the results of the SED fitting, this assumption holds even in the galactic centre except for the AGN \citep{Saito2016}.
\citet{Miyamoto2017} showed that the spectral index derived from 4.9 GHz and 95 GHz continuum images at the ring is $\sim -0.2$, indicating that the 110~GHz continuum is primarily dominated by the free-free emission from young star-forming regions, and hence the 110 GHz continuum can be a good tracer of SFR.
As the main target in this study is the star-forming ring, we can ignore the nonthermal emission from the AGN in estimating SFR.
The continuum data were retrieved from 2015.1.01487.S, which consists of two 12 m array configurations with five spectral windows for scientific observations.
We processed the data using the observatory-provided calibration scripts on CASA \citep{CASA2022} version 4.7.2. 
After inspecting the calibrated data, the continuum emission was identified, and the contribution of the lines was subtracted for each visibility datapoint of the array configurations and spectral windows.
The imaging of the continuum was performed using \texttt{tclean} in CASA version 6.1.0 via multifrequency synthesis using five spectral windows.
We applied tapering to the data in baselines that were longer than 420 k$\lambda$ and a Briggs weighting with a robust parameter of 0.5, which resulted in a synthesized beam of $0\farcs50\times0\farcs49$ (PA = $-21\fdg$94).
The final data were spatially smoothed to obtain an angular resolution of 0\farcs8 with a grid of 0$\farcs$4. 
The rms noise of the resultant 110~GHz continuum was 14 $\mu$Jy beam$^{-1}$.
Figure~\ref{fig:110GHz} shows the 110~GHz continuum image smoothed to $0\farcs8$ resolution.

\begin{figure}
	\begin{center}
		\includegraphics[width=80mm]{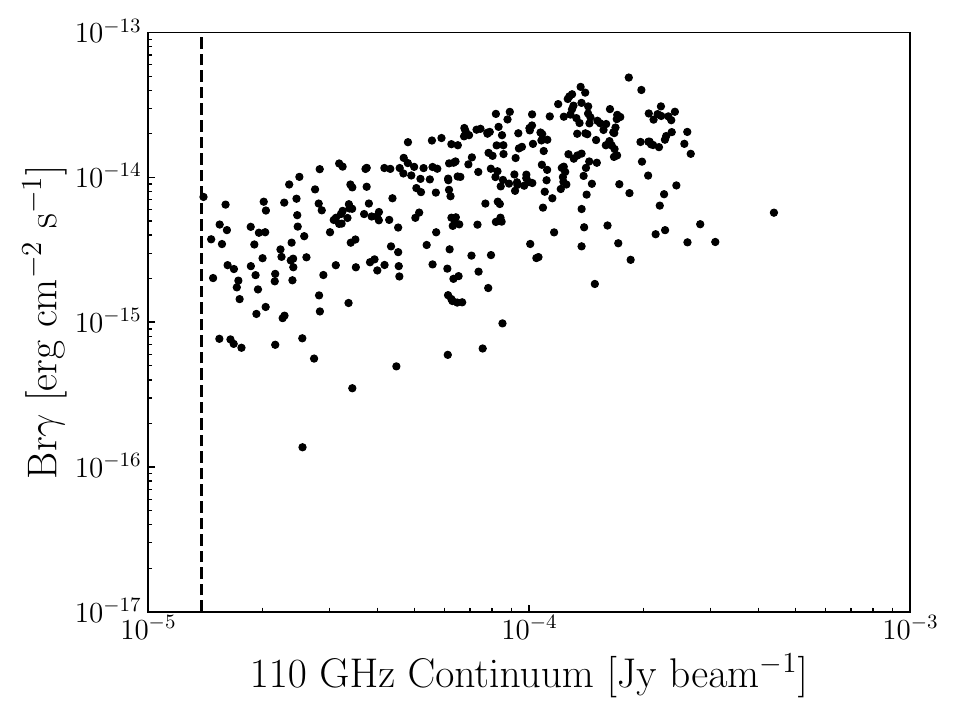}
	\end{center}
	\caption{Relationship between the 110~GHz continuum and Br$\gamma$ emission. The vertical dashed line represents the 1-$\sigma$ sensitivity limit of the 110~GHz continuum (1$\sigma = 1.38 \times 10^{-5}$ Jy beam$^{-1}$).}
	\label{fig:100GHz-BrGplot}
\end{figure}

Figure~\ref{fig:110GHz} compares the spatial distributions of Br$\gamma$ \citep[contours;][]{Falcon-Barroso2014} and the 110 GHz continuum image (color map).
Their emission peaks approximately coincide; spatial displacement between Br$\gamma$ and 110 GHz continuum peaks is at most 2$^{\prime\prime}$ -- 3$^{\prime\prime}$ ($\sim$200~pc).
However, the Br$\gamma$ to 110 GHz intensity ratio substantially varies from peak to peak.
For example, the southernmost Br$\gamma$ peak is not prominent at 110~GHz, whereas the eastern peaks are brighter at 110~GHz. 
Moreover, the southeastern Br$\gamma$ peak, which is relatively strong, is considerably weak at 110~GHz.
A similar trend is identified in the pixel-by-pixel scatter plot of Br$\gamma$ and 110 GHz continuum shown in figure~\ref{fig:100GHz-BrGplot}; the plot shows a significantly large scatter, apparently upper-limited by the linear relationship line.
This variation in the Br$\gamma$-to-110 GHz ratio is likely due to the extinction of Br$\gamma$ by dust.
Indeed, the 350 GHz dust emission \citep{Miyamoto2017} is bright at weak Br$\gamma$ peaks, while it is weak around bright Br$\gamma$ peaks in the southeastern region. 
These results suggest that some star-forming regions in the central kpc region of NGC\,613 are dust-embedded, and Br$\gamma$ cannot accurately trace SFR in such regions.

We convert the 110 GHz continuum flux into SFR using the equation \citep{Yun2002}:
\begin{equation}
	S = 0.71\nu^{-0.1}\frac{\mathrm{SFR}}{M_\sun \ \mathrm{yr}^{-1}}D_{L}^{-2}(1+\mathrm{z}),
\end{equation}
where {\itshape S}, $\nu$, $D_{L}$, and z are the continuum flux in Jy, the rest frequency in GHz, luminosity distance in Mpc, and redshift, respectively.
In the subsequent analysis, we mask the central 1$\farcs$45 region since the 110 GHz emission is likely contaminated by the AGN emission.

\subsection{Method\label{subsection:Method:HBI}} 
\citet{Sato2021} showed that multiple velocity components spatially overlap in the position-velocity diagram (PVD) of the central region of NGC\,613, indicating that the molecular gas is colliding.
We show a PVD of the HCN($J$ = 1--0)/$^{13}$CO($J$ = 1--0) ratio in Appendix \ref{sec:appendix_line_ratio}, which reveals that the ratio varies along the velocity direction.
Two-dimensional analysis based on the integrated intensity map is unable to distinguish these multiple velocity components with different line intensity ratios.
Therefore, we performed 3D analysis using the 3D data cubes to accurately estimate the physical properties of molecular clouds.

\begin{deluxetable*}{lll}
	\tablecaption{Model Parameters\label{tab:modelparam}}
	\tablehead{\colhead{Parameter} & \colhead{Unit} & \colhead{Description}}
	\startdata
	$\log_{10}$\ d$N$/d$v$ & cm$^{-2}$\,(km\,s$^{-1}$)$^{-1}$ & Molecular hydrogen column density per velocity width \\
	$\log_{10}\ n_\mathrm{H_2}$ & cm$^{-3}$ & Molecular hydrogen volume density \\                    
	$\log_{10}\ T_\mathrm{kin}$ & K & Gas kinetic temperature \\  
	$\log_{10}\ \phi$ & & Parameter defining the product of the beam-filling factor \\
	~ & ~ & and the velocity channel filling factor $\Phi$ ($\Phi \equiv 1 - e^{-\phi}$) \\
	$\log_{10}\ \xmolp{^{13}\mathrm{CO}}^{*}$ & & Fractional molecular abundance to $\mathrm{H_2}$\\
	$\log_{10}\ \xmolp{\mathrm{C^{18}O}}$ & & \\ 
	$\log_{10}\ \xmolp{\mathrm{CS}}$ & & \\
	$\log_{10}\ \xmolp{\mathrm{HCN}}$ & & \\
	$\log_{10}\ \xmolp{\mathrm{HCO^{+}}}$ & & \\
	\enddata
	\tablecomments{$^{13}$CO fractional abundance $\xmolp{^{13}CO}$ is assumed to be constant at $10^{-5.5}$ in the model.}
\end{deluxetable*}

We calculated PPV distributions of the molecular hydrogen column density $N_\mathrm{H_2}$, $n_\mathrm{H_2}$, and $T_\mathrm{kin}$ by solving non-LTE excitation equations based on the lines in table \ref{tab:spe_para}.
The radiative and collisional coefficients were obtained from the Leiden Atomic and Molecular Data Base \citep[LAMDA;][]{Schoier2005}.
The large velocity gradient (LVG) approximation \citep{Goldreich1974} was used to calculate the photon-trapping effect.
All input line intensities were convolved to an angular resolution of $0\farcs$8 to match the 110~GHz continuum, the lowest resolution among the data used in this study.
They were then resampled to a $0\farcs4\times0\farcs4\times10\ \mathrm{km\,s}^{-1}$ PPV grid in order to be sampled at the Nyquist sampling rate and converted into a brightness temperature scale.
The free parameters in the model are listed in table \ref{tab:modelparam}: d$N$/d$v$, which is molecular hydrogen column density $N$ per velocity width $v$, $n_\mathrm{H_2}$, $T_\mathrm{kin}$, $\phi$ ($\equiv -\ln (1-\Phi)$), where $\Phi$ is the product of the beam-filling factor and the velocity channel filling factor, and the molecular abundances relative to $\mathrm{H_2}$ ($\xmolp{H_2}$).
All parameters are defined on a base-10 log scale during the calculations.
We approximate that the $^{13}\mathrm{CO}$ fractional abundance $\xmolp{^{13}CO}$ is constant throughout the analysis region to use $^{13}\mathrm{CO}$ as a proxy for $\mathrm{H_2}$.
Since no accurate measurement of $\xmolp{^{13}CO}$ has been reported for NGC\,613, we assume $\xmolp{^{13}CO} = 10^{-5.5}$, which is a typical value of the Perseus molecular clouds \citep[$10^{-5.4}$--$10^{-5.7}$;][]{Pineda2008}.
Similar values were reported for the central regions of the Galaxy ($10^{-5.4}$; e.g., T18) and NGC\,253 \citep[$10^{-5.3}$;][]{Martin2019}.
We note a caveat that what we calculate as the column density $N_\mathrm{H_2}$ in this paper is actually d$N$/d$v$ 10$^{5.5}\times\frac{[\mathrm{^{13}CO}]}{[\mathrm{H_2}]}$, where $\frac{[\mathrm{^{13}CO}]}{[\mathrm{H_2}]}$ is the true fractional abundance of $^{13}\mathrm{CO}$.
Hence, the uncertainty in $\frac{[\mathrm{^{13}CO}]}{[\mathrm{H_2}]}$ and its spatial variation could add uncertainty to the H$_{2}$ column density and molecular gas mass.
The calculations were performed on the voxels, where both the $^{13}$CO(1--0) and C$^{18}$O(1--0) fluxes were larger than 2$\sigma$.
This initial filtering using a signal-to-noise ratio excludes low column-density molecular gas.
We note that such low column-density gas can contribute to the total molecular gas mass, which may link to star formation activity.

The HB inference framework used in T18 was employed to solve the excitation equations.
The HB analysis can handle nonstatistical errors in the input line intensities, owing to calibration uncertainties and deficiencies of the model, which are difficult to treat with standard maximum-likelihood analysis.  
The basis of the HB method is provided in T18 and described in Appendix \ref{sec:appendix_HB}.
Here, we outline the modifications from the T18 analysis.
The details of the modifications are also written in Appendix \ref{sec:appendix_HB}.

We made two modifications to the method developed by T18.
First, we used two additional logistic hyperpriors not to make the prior correlation coefficients for ($N_\mathrm{H_2}$, $T_\mathrm{kin}$) and ($T_\mathrm{kin}$, $n_\mathrm{H_2}$) negative.
This modification inhibits the fast degeneration of $N_\mathrm{H_2}$, $n_\mathrm{H_2}$, and $T_\mathrm{kin}$ in the likelihood function, caused by the smaller number of input lines than that in T18.
They may mask true anticorrelation present in the real clouds.
However, some observational results suggest the absence of strong anticorrelation of $T_\mathrm{kin}$ with $N_\mathrm{H_2}$ (and hence with $n_\mathrm{H_2}$) in dense interstellar medium (ISM).
For example, T18 found that the $T_\mathrm{kin}$ distribution has a zero or slightly positive correlation with $N_\mathrm{H_2}$. 
This result is consistent with \citet{Ott2014} and \citet{Ginsburg2016}, in which high $T_\mathrm{kin}$ regions often coincide with $N_\mathrm{H_2}$ peaks.
Star formation generally occurs in high-density regions under strong turbulence. 
Since the stellar feedback makes the surrounding temperature higher in such regions, we may expect a positive correlation between density and temperature.
Second, we modified logistic hyperpriors to set lower limits on its elements.
The original logistic hyperpriors introduced in T18 were only set upper limits.
Therefore, our model can put both lower and upper limits on the elements.
These modifications make the HB analysis solve the excitation equations for our dataset within a reasonable computational time.

The non-LTE method with HB inference we used has several advantages over previous non-LTE analyzes.
HB inference can suppress the artificial correlations among the parameters created by systematic errors owing to a deficiency in the simple one-zone excitation analysis and calibration uncertainty (T18).
We compared this advantage with the previous non-LTE analysis in Appendix \ref{sec:appendix_comparison}.
Additionally, derived maps do not show implausible anticorrelations in distributions and outliers when input data are maps.
These advantages are suitable for deriving gas properties with high-resolution multiline imaging data.

The present study assumes a common $\Phi$ for all lines at each voxel.
This may be a relevant assumption to investigate the physical conditions of dense gas since higher-density tracers (e.g. band-7 lines of HCN, HCO$^+$, and CS) should actually have lower effective beam-filling factors than low-$J$ CO isotopologueue lines.  
However, the common-$\Phi$ assumption allows us to estimate ``mean'' $n_\mathrm{H_2}$ over a 0$\farcs$8 beam area; assuming a single $\Phi$ value irrespective of actual source sizes of different density media is equivalent to estimating the representative physical properties of molecular gas confined in the beam. 

\section{Results}\label{sec:results}
\subsection{Three-dimensional view of molecular cloud properties in the central region of NGC 613\label{subsection:Results:3Dviews}}

\begin{figure*}[htp]
	\begin{center}
		\includegraphics[width=160mm]{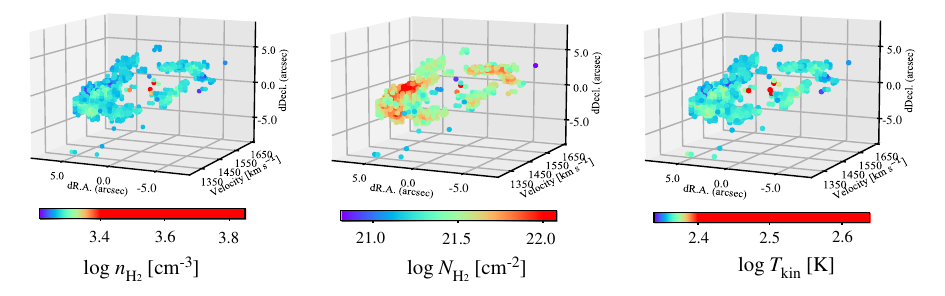} 
	\end{center}
	\caption{Three-dimensional (PPV) distributions of molecular hydrogen column density ($N_\mathrm{H_2}$), volume density ($n_\mathrm{H_2}$), and gas kinetic temperature ($T_\mathrm{kin}$) in the central $20'' \times 20''$ (1.7 kpc $\times$ 1.7 kpc) region of NGC\,613 (from left to right). 
		The velocity axis ranges from $V_\mathrm{LSR}$ = 1250--1650 km s$^{-1}$.}\label{fig:nNH2T}
\end{figure*}

\begin{figure*}[htp]
	\begin{center}
		\includegraphics[width=160mm]{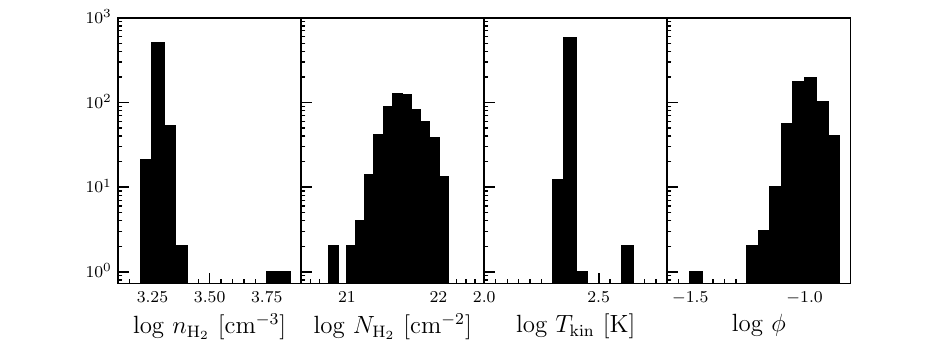} 
	\end{center}
	\caption{Histograms of molecular hydrogen volume density $n_\mathrm{H_2}$, column density $N_\mathrm{H_2}$, kinetic temperature $T_\mathrm{kin}$, and the parameter defining beam-filling factor $\phi$. These values are derived on a voxel-to-voxel basis.}\label{fig:nNH2T_histo}
\end{figure*}

The PPV views of $n_\mathrm{H_2}$, $N_\mathrm{H_2}$, and $T_\mathrm{kin}$ calculated by the HB method are shown in figure~\ref{fig:nNH2T}.
The histograms of these quantities and $\phi$ are displayed in figure~\ref{fig:nNH2T_histo}.
We define the CND as a central 1$\farcs$45 region (a radius of 124 pc) and the star-forming ring as the rest of the calculated regions.
Table~\ref{tab:second} shows the means and standard deviations of the physical properties of the GMCs in the CND and the star-forming ring;
note that they are voxel-based values, not their line-of-sight averages.

The histograms (figure \ref{fig:nNH2T_histo}) show that $N_\mathrm{H2}$ ranges from $6.4\times10^{20}$ to 1.2$\times$10$^{22}$ cm$^{-2}$ (10$^{20.8-22.1}$ cm$^{-2}$).
While $n_\mathrm{H_2}$ ranges from $1.6\times10^{3}$ to $7.1\times10^{3}$ cm$^{-3}$ ($10^{3.21-3.85}$ cm$^{-3}$), most regions show a constant density of $\sim10^{3.2-3.4}$ cm$^{-3}$.
The $T_\mathrm{kin}$ distribution has a range of 220--440~K ($10^{2.33-2.64}$~K), most of which are lower than 250~K.
As seen in figure~\ref{fig:nNH2T_histo}, $n_\mathrm{H_2}$ values are within an order of magnitude except for a few voxels corresponding to the CND, whereas $N_\mathrm{H_2}$ spans 2 orders of magnitude.
The $\phi$ values in the star-forming ring have $0.11\pm0.01$, while those in the CND are $0.08\pm0.02$.

We compare the derived gas properties between the CND and star-forming ring.
The $n_\mathrm{H_2}$ and $T_\mathrm{kin}$ values are higher in the CND than in the ring, whereas $N_\mathrm{H_2}$ is lower in the CND.
This tendency can also see in 2D distributions, as discussed in subsection \ref{subsec:2D}.
The ring exhibits a higher $N_\mathrm{H_2}$ on the eastern side than on the western side.
The $T_\mathrm{kin}$ value is approximately constant in the ring at $\sim$2.3~$\times$ 10$^{2}$~K (10$^{2.36}$~K).
We obtained an almost uniform $n_\mathrm{H_2}$ of $\sim$1.9$\times$10$^{3}$~cm$^{-3}$ (10$^{3.28}$~cm$^{-3}$) across the ring.

The $n_\mathrm{H_2}$ value of the star-forming ring calculated in the present study is lower than that of \citet{Miyamoto2017} ($n_\mathrm{H_2}$ = 10$^{4-4.5}$~cm$^{-3}$) and comparable to or slightly higher than that reported by \citet{Sato2021} ($n_\mathrm{H_2}$ = 10$^{2.6-3.8}$~cm$^{-3}$).
This difference may be explained with three possibilities.
The first possibility is the different estimation processes.
\citet{Miyamoto2017} estimated the $n_\mathrm{H_2}$ value after averaging the line intensities over the ring.
In contrast, the present study and \citet{Sato2021} derived physical quantities leaving the data points spatially resolved.
As discussed in Appendix \ref{sec:appendix_line_ratio}, the input line ratio varies within the ring.
Therefore, averaging the line ratios over the ring would not represent the physical conditions of each position in the ring.
The second possibility is the different tracer-line selections.
\citet{Miyamoto2017} used multi$J$ lines of HCN, HCO$^{+}$, and CS to calculate the physical conditions.
Thus, their results are biased toward higher $T_\mathrm{kin}$ and/or $n_\mathrm{H_2}$ regions than ours because our study also used CO and its isotopologue lines, which trace lower $T_\mathrm{kin}$ and $n_\mathrm{H_2}$ regions.
On the other hand, \citet{Sato2021} used only $J\leq3$ CO and isotopologue lines in their excitation analyzes; this should cause a bias toward a medium with lower $n_\mathrm{H_2}$ and/or $T_\mathrm{kin}$ than that we analyzed.
The third possibility is methodology.
Our method, which uses hierarchical Bayesian inference and is derived in three dimensions, differs from what they used (RADEX modeling in two dimensions).
However, since the difference between their results and ours is within 1 order of magnitude, it is considered that our method can at least estimate the volume density with the reliability of a factor of few.
In addition, the result of our method that the volume density is almost constant within the star-forming ring is consistent with \citet{Sato2021} that there is no volume density gradient in the star-forming ring except in some regions connecting with the bar structure.
This fact also suggests that although we introduced the nonnegative prior correlation coefficients for ($N_\mathrm{H_2}$, $T_\mathrm{kin}$) and ($T_\mathrm{kin}$, $n_\mathrm{H_2}$), we could estimate the typical values for $T_\mathrm{kin}$ and $n_\mathrm{H_2}$.
We conclude that it should be reasonable that the obtained $n_\mathrm{H_2}$ value is between those in the two previous analyzes.

\begin{deluxetable*}{cccccc}
	\tablecaption{Means and dispersions of the physical quantities\label{tab:second}}
	\tablehead{
		\colhead{Region} & \colhead{Voxels} & $n_\mathrm{H_2}$ [cm$^{-3}$] & $N_\mathrm{H_2}$ [cm$^{-2}$] & $T_\mathrm{kin}$ [K] & $\phi$
	}
	\startdata
	All & 587 & $(1.9\pm0.3)\times10^{3}$ & $(4.6\pm2.0)\times10^{21}$ & $(2.3\pm0.1)\times10^{2}$ &  $0.10\pm0.01$\\
	Circumnuclear Disk & 12 & $(2.7\pm1.7)\times10^{3}$ & $(3.0\pm3.1)\times10^{21}$ & $(2.7\pm0.7)\times10^{2}$ & $0.08\pm0.02$\\
	Star-forming Ring & 575 & $(1.9\pm0.1)\times10^{3}$ & $(4.6\pm2.0)\times10^{21}$ & $(2.3\pm0.1)\times10^{2}$  & $0.11\pm0.01$\\
	\enddata
\end{deluxetable*}

\subsection{Two-dimensional view of physical quantities of molecular clouds}\label{subsec:2D}
\begin{figure*}[htp]
	\begin{center}
		\includegraphics[width=160mm]{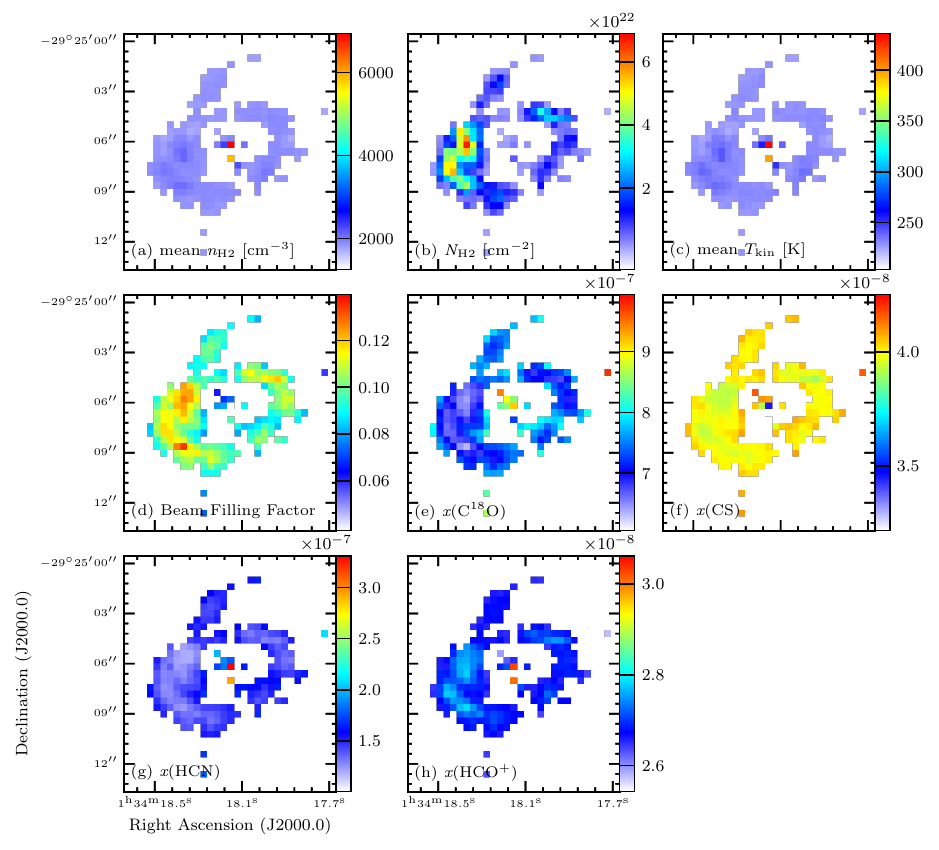}
	\end{center}
	\caption{Two-dimensional maps of derived quantities. (a): mean H$_2$ volume density, (b): H$_2$ column density integrated over the velocity axis after the beam-filling factor weighting, (c): mean kinematic temperature, (d): mean beam-filling factor, (e) mean C$^{18}$O fractional abundance, (f) mean CS fractional abundance, (g) mean HCN fractional abundance, and (h) mean HCO$^+$ fractional abundance. Except for (b), we derived the values after weighting with H$_2$-column density and beam-filling factor for each voxel.}
	\label{fig:2D}
\end{figure*}

To generate the position-position 2D distribution of $N_\mathrm{H_2}$, we first multiplied the beam-filling factor $\phi$ and $N_\mathrm{H_2}$ cubes in each channel to correct for $\phi$.
Next, we integrated this cube in the velocity direction.
Similarly, we created 2D maps of $n_\mathrm{H_2}$ and $T_\mathrm{kin}$.
These were weighted by $N_\mathrm{H_2}\times \phi$ and then averaged in the velocity direction.
The 2D maps of $N_\mathrm{H_2}$, $n_\mathrm{H_2}$, and $T_\mathrm{kin}$ are shown in figures \ref{fig:2D} (a), (b), and (c), respectively.
The $n_\mathrm{H_2}$ and $T_\mathrm{kin}$ are higher in the CND than in the star-forming ring, which is more evident than in the 3D map.

Figures \ref{fig:2D} (d)--(h) show the 2D maps of the beam-filling factor and the fractional abundance of C$^{18}$O, CS, HCN, and HCO$^+$, which are obtained in the same manner as the 2D $n_\mathrm{H_2}$ and $T_\mathrm{kin}$ maps.
The beam-filling factor is slightly higher on the eastern side of the star-forming ring than that on the western side.
The fractional abundance maps show that the CND has higher abundances than the star-forming ring in all molecules.
We find a spatial difference in the molecular abundances between the eastern and western sides of the star-forming ring;
the eastern side has higher CS and HCO$^{+}$ abundances but lower C$^{18}$O and HCN abundances than the western side.
The fractional abundance CS, HCN, and HCO$^{+}$ is consistent by a factor of 3 with \citet{Usero2004}, who determined the molecular abundances in the star-forming ring of the AGN, NGC 1068, using the large velocity gradient (LVG) analysis.
This fact suggests the fractional abundance derived from the HB method is compatible with those from the typical modelling, at least the LVG model.

\subsection{Properties of molecular gas and star formation}\label{subsec:GasProperties}

\begin{figure}
	\begin{center}
		\includegraphics[width=80mm]{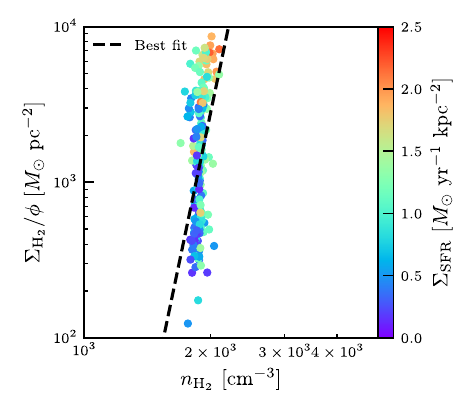} 
	\end{center}
	\caption{Relationship between volume density $n_\mathrm{H_2}$ and the surface density divided by the beam-filling factor $\Sigma_{\mathrm{H_2}}/\phi$. 
		The color of each data point indicates the surface density of SFR $\Sigma_{\mathrm{SFR}}$. 
		The dashed line shows the best fit: log$_{10} \ \Sigma_\mathrm{H_2}/\phi $ = 12.9 log$_{10} \ n_\mathrm{H_2}$ -39.3, although this relationship might be nonphysical due to a narrow dynamic range of $n_\mathrm{H_2}$.}\label{fig:NH2-n}
\end{figure}

\begin{figure}
	\begin{center}
		\includegraphics[width=80mm]{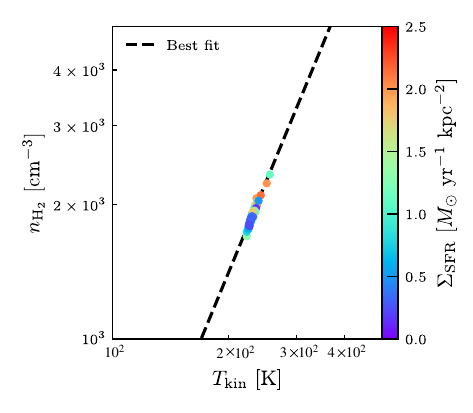} 
	\end{center}
	\caption{Relationship between kinetic temperature $T_\mathrm{kin}$ and volume density $n_\mathrm{H_2}$. 
		The colour scale is the same as that in figure~\ref{fig:NH2-n}. 
		The dashed line indicates the best fit: log$_{10} \ T_\mathrm{kin}$ = 2.09 log$_{10} \ n_\mathrm{H_2}$ -1.66, although this relationship might be nonphysical due to the hyperprior that ($T_\mathrm{kin}$, $n_\mathrm{H_2}$) is not negatively correlated.}\label{fig:n-T}
\end{figure}

In this subsection, we investigate the relationship among the derived physical quantities.
In the subsequent analysis, we use the molecular hydrogen surface density $\Sigma_\mathrm{H_2}$ instead of $N_\mathrm{H_2}$ for the convenience of comparison with previous studies.
Figures \ref{fig:NH2-n} and \ref{fig:n-T} show scatter plots of $n_\mathrm{H_2}$-$\Sigma_\mathrm{H_2}$/$\phi$ and $T_\mathrm{kin}$-$n_\mathrm{H_2}$.
As described in subsection \ref{subsec:2D}, the 2D $N_\mathrm{H_2}$ is corrected for the beam-filling factor.
Therefore, $\Sigma_{\mathrm{H_2}}$ is suitable for comparison with $\Sigma_{\mathrm{SFR}}$ (which will be discussed later) because it reflects the distributions of molecular gas.
However, to investigate the intrinsic properties of molecular gas ($n_\mathrm{H_2}$ and $T_\mathrm{kin}$), it is better to see relationships to $\Sigma_{\mathrm{H_2}}/\phi$.
The data points are colored according to $\Sigma_\mathrm{SFR}$, and the best-fit line is shown as a dashed line.
The typical error of each datum is less than 15\%.
We do not use the CND voxels in the following analysis since they are not of immediate interest to this study.

Figure~\ref{fig:NH2-n} does not show a clear relationship between $n_\mathrm{H_2}$ and $\Sigma_\mathrm{H_2}/\phi$, which is due to a narrow dynamic range of $n_\mathrm{H_2}$ obtained in our inference, though a weak positive correlation is observed in $\Sigma_\mathrm{H_2}/\phi \gtrsim 4\times10^{3}$ $M_\odot$ pc$^{-2}$.
Figure~\ref{fig:n-T} illustrates that $T_\mathrm{kin}$ has a tight correlation with $n_\mathrm{H_2}$.
As seen in subsection \ref{subsection:Results:3Dviews}, the typical values for $T_\mathrm{kin}$ and $n_\mathrm{H_2}$ do not differ significantly from the previous results, meaning that the overall values would not be unrealistic.
However, this result can be attributed to the hyperprior that ($T_\mathrm{kin}$, $n_\mathrm{H_2}$) is not negatively correlated, as written in subsection \ref{subsection:Method:HBI}, and thus this relation does not seem to be a true correlation in molecular clouds, and the degeneracy between $T_\mathrm{kin}$ and $n_\mathrm{H_2}$ is not fully solved.
Comparing figures \ref{fig:maps} and \ref{fig:2D}, the distributions of the line intensity are not similar to those of  $T_\mathrm{kin}$ and $n_\mathrm{H_2}$, but similar to that of $N_\mathrm{H_2}$ ($\Sigma_\mathrm{H_2}$). 
Considering the poor correlation between $n_\mathrm{H_2}$ and $\Sigma_\mathrm{H_2}/\phi$, the line intensities do not seem to make a significant contribution to solving the degeneracy between $T_\mathrm{kin}$ and $n_\mathrm{H_2}$. On the other hand, the excitation temperature is strongly dependent on the line intensities. 
The excitation temperature is a quantity defined from the ratio of column densities at a given two levels. 
Thus, if the line ratios have weak variation, the line-to-mass conversion factor could be relatively constant. 
In such a case, $N_\mathrm{H_2}$ and the integrated intensity maps would show similar distributions as in this inference. 
We conclude that the estimated $N_\mathrm{H_2}$ is reliable even if the relative contributions from $T_\mathrm{kin}$ and $n_\mathrm{H_2}$ to excitation are not well solved.

\begin{figure}
	\begin{center}
		\includegraphics[width=80mm]{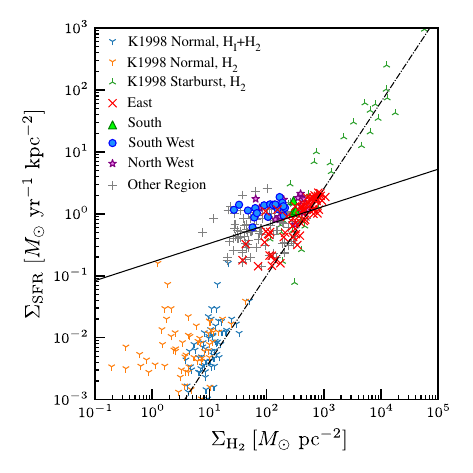} 
	\end{center}
	\caption{Relationship between the surface density of molecular hydrogen gas $\Sigma_{\mathrm{H_2}}$ and SFR $\Sigma_{\mathrm{SFR}}$ in the star-forming ring of NGC\,613, compared with nearby galaxies in a galaxy scale \citep{Kennicutt1998}. 
		Eastern, southern, northwestern, and southwestern subregions are represented by red crosses, green triangles, purple asterisks, and blue circles, respectively. 
		The solid line and dashed-dotted line indicate the best fit for the star-forming ring of NGC\,613: log$_{10}\Sigma_\mathrm{SFR}$ = 0.30 log$_{10}\Sigma_\mathrm{H_{2}}$ -0.78 and the galaxy-scale Kennicutt--Schmidt law in nearby star-forming galaxies, respectively \citep{Kennicutt1998}.}\label{fig:KS-law}
\end{figure}

Figure~\ref{fig:KS-law} presents a scatter plot of the surface density of molecular hydrogen gas $\Sigma_{\mathrm{H_2}}$ and SFR $\Sigma_{\mathrm{SFR}}$ in the star-forming ring of NGC\,613, overlaid on the galaxy-scale Kennicutt--Schmidt (KS) law \citep{Kennicutt1998} of nearby star-forming galaxies.
The subregions of the star-forming ring of NGC\,613 defined in Figure \ref{fig:maps} are represented by different symbols.
Since we use different $\Sigma_\mathrm{H_2}$ and $\Sigma_\mathrm{SFR}$ tracers than those used by \citet{Kennicutt1998}, the index and scatter of the $\Sigma_\mathrm{H_2}$-$\Sigma_\mathrm{SFR}$ relationship may vary from theirs.
\citet{Leroy2013} considered the effect on the index and scatter of the $\Sigma_\mathrm{H_2}$-$\Sigma_\mathrm{SFR}$ relationship with different $\Sigma_\mathrm{H_2}$ and $\Sigma_\mathrm{SFR}$ tracers at the kpc scale.
Note, however, that the SFR tracers discussed by \citet{Leroy2013} did not include the 110 GHz continuum used in this study.
Although there is some variation in the index in individual galaxies (almost within 0.2), changing tracers have a mild impact on the overall index and scatter.
They also examined the effects of different spatial resolutions through comparison with the literature.
The change in star formation efficiency (SFE; $\Sigma_{\mathrm{SFR}}/\Sigma_{\mathrm{H_2}}$) due to different resolutions is about a factor of 2.
Its scatter is slightly higher on the high-resolution case, but generally below 0.4 dex.

The $\Sigma_{\mathrm{H_2}}-\Sigma_{\mathrm{SFR}}$ in the star-forming ring of NGC\,613 shows an overall positive correlation.
However, its index of 0.30 (solid line) with a large scatter ($\sim$1 dex) is significantly smaller than the $\Sigma_{\mathrm{H_2}}-\Sigma_{\mathrm{SFR}}$ on the galaxy scale \citep[1.4, dashed-dotted line: ][]{Kennicutt1998} and $\sim$100 pc scale \citep[$\sim$1.0: ][]{Pessa2021}.
The molecular clouds in the star-forming ring of NGC\,613 exhibit higher SFRs than the galaxy-scale KS law (dashed-dotted line), indicative of high SFEs.
The $\Sigma_\mathrm{H_2}$ value of NGC\,613 ranges from normal galaxy values to starburst values in \citet{Kennicutt1998}.
The $\Sigma_\mathrm{H_2}$--$\Sigma_\mathrm{SFR}$ relation in NGC\,613 is 1-2 orders of magnitude above the galaxy-scale KS law in the $\Sigma_\mathrm{H_2}$ range of normal galaxies and appears to asymptote toward the galaxy-scale KS law at the highest $\Sigma_\mathrm{H_2}$ end.
There are two sequences with different power-law indices in the plot; the southwestern and a part of northwestern clouds tend to have generally elevated $\Sigma_\mathrm{SFR}$ values with no $\Sigma_\mathrm{H_2}$-dependence, whereas the other subregions appear to follow an approximately linear relationship.
The SFE in the southwestern and a part of northwestern clouds is $\sim$0.5 dex higher than in the other subregions.
This value is larger than a scatter of SFE in nearby galaxies at the $\sim$100 pc scale \citep[$\sim$0.4 dex][]{Pessa2021}, suggesting that the mechanism of star formation at least in the southwestern subregion is different from other subregions.

\begin{figure}
	\begin{center}
		\includegraphics[width=80mm]{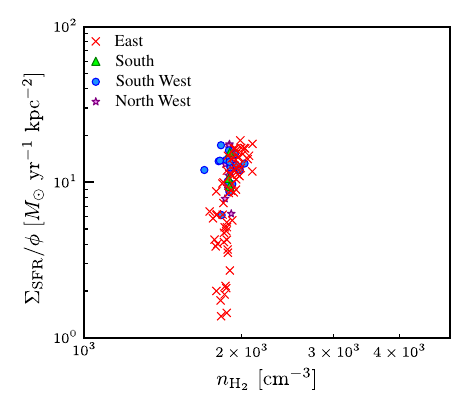} 
	\end{center}
	\caption{Relationship between volume density $n_\mathrm{H_2}$ and the surface density of SFR divided by the beam-filling factor $\Sigma_\mathrm{SFR}/\phi$ for each subregion.}\label{fig:n-SFR}
\end{figure}

Figure~\ref{fig:n-SFR} displays the scatter plot between $n_\mathrm{H_2}$ and $\Sigma_{\mathrm{SFR}}/\phi$.
Since $\Sigma_{\mathrm{SFR}}$ is a beam-averaged value, a physical comparison with $n_\mathrm{H_2}$ can be made by dividing $\Sigma_{\mathrm{SFR}}$ by the beam-filling factor.
As figure~\ref{fig:KS-law}, we plot them with different symbols for each subregion, but the molecular clouds in the other region are not shown for clarity.
Due to the narrow dynamic range of $n_\mathrm{H_2}$, there is no difference between the subregions, given that the estimation accuracy is $\lesssim$15\%.

\section{Discussion}\label{sec:discuss}
\subsection{Origin of the Scatter in the Surface Density of H$_{2}$ and SFR Diagram}
The important results from the physical condition measurements we made with the ALMA archival data are summarized as follows:
\begin{itemize}
	\item $n_\mathrm{H_2}$ and $T_\mathrm{kin}$ are mostly homogeneous in the star-forming ring.
	\item The $\Sigma_{\mathrm{H_2}}$--$\Sigma_\mathrm{SFR}$ diagram in the star-forming ring exhibits a large scatter ($\sim$1 dex) and follows a power-law relation with an index of 0.3, which is significantly lower than that in nearby galaxies with a similar resolution ($\sim$1.0).
\end{itemize}
In the following, we investigate the environmental dependence of star formation properties of the NGC\,613 star-forming ring based on these results.

The mean values of the physical parameters and their dispersions within each region are summarized in table \ref{tab:each_area}. 
All subregions have similar mean $T_\mathrm{kin}$ and $n_\mathrm{H_2}$, but the southwestern region has lower $\Sigma_\mathrm{H_2}$ than other subregions.

\begin{deluxetable*}{ccccc}
	\tablecaption{Means and dispersions of the physical quantities in each region\label{tab:each_area}}
	\tablehead{
		Region & Pixels &  $n_\mathrm{H2}$ [cm$^{-3}$]  & $\Sigma_\mathrm{H_2}$ [$M_\sun$~pc$^{-2}$]  &  $T_\mathrm{kin}$ [K]}
	\decimals
	\startdata
	Ring & 212 & $(1.9\pm0.1)\times$10$^{3}$ & $(2.2\pm2.2)\times10^{2}$ & $(2.3\pm0.1)\times$10$^{2}$\\
	East & 68 & $(1.9\pm0.1)\times$10$^{3}$ & $(4.2\pm2.7)\times10^{2}$ & $(2.3\pm0.1)\times$10$^{2}$\\
	South & 3 & $(1.9\pm0.1)\times$10$^{3}$ & $(1.9\pm0.1)\times10^{2}$ & $(2.3\pm0.1)\times$10$^{2}$\\
	Southwest & 21 & $(1.9\pm0.1)\times$10$^{3}$ & $(1.2\pm0.6)\times10^{2}$ & $(2.3\pm0.1)\times$10$^{2}$\\
	Northwest & 14 & $(1.9\pm0.1)\times$10$^{3}$ & $(2.8\pm1.5)\times10^{2}$ & $(2.3\pm0.1)\times$10$^{3}$\\
	\enddata
\end{deluxetable*}

\begin{figure}
	\begin{center}
		\includegraphics[width=80mm]{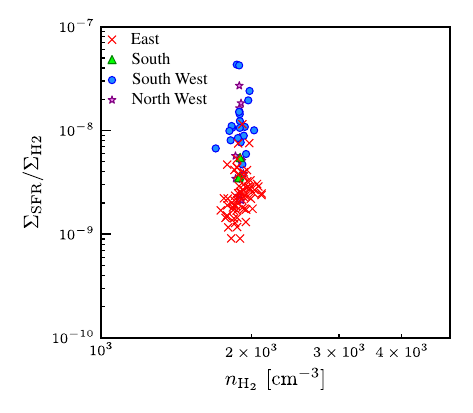}
	\end{center}
	\caption{Relationship between mean volume density and $\Sigma_\mathrm{SFR}/\Sigma_\mathrm{H_2}$. The symbols are the same as in figure~\ref{fig:n-SFR}.} \label{fig:n-SFE}
\end{figure}

As described in subsection \ref{subsec:GasProperties}, the large scatter in the low-$\Sigma_\mathrm{H_2}$ region in the $\Sigma_\mathrm{H_2}$-$\Sigma_\mathrm{SFR}$ diagram originates from the elevated SFE in the southwestern subregion.
The elevated SFE in the southwestern subregion may suggest a star formation driver working specifically there.
Enhancement in $n_\mathrm{H2}$ is a candidate for a driver of such SFE variation \citep[e.g.,][]{GaoSolomon2004,Muraoka2009,Usero2015,Bigiel2016,Yajima2019}.  
However, $n_\mathrm{H_2}$ does not systematically differ among subregions, as can be seen in figure~\ref{fig:n-SFE} and table \ref{tab:each_area}, despite their different SFEs.
We again note that this result should not be due to the limitation of the hyperpriors we used, since the non-LTE calculations by \citet{Sato2021} also showed no density gradient in the star-forming ring.
The absence of the effect of $n_\mathrm{H_2}$ on the star formation in the star-forming ring is also seen in the $n_\mathrm{H_2}$--$\Sigma_\mathrm{SFR}/\phi$ relationship (figure~\ref{fig:n-SFR}).
Although $n_\mathrm{H_2}$ is distributed in a narrow range, the two sequence patterns (i.e., the southwestern and northwestern subregions and the others) are not visible in the $n_\mathrm{H_2}$--$\Sigma_\mathrm{SFR}/\phi$ plane.
These results imply that $n_\mathrm{H2}$ is not the key parameter to govern the SFE variations in the ring of NGC\,613. 

We may consider two possible causes.
The first is efficient star formation triggers that work specifically in the southwestern and northwestern subregions, such as rapid gas infall or cloud--cloud collisions. 
The second is a deficiency of molecular gas due to a high gas consumption rate or lack of supply.

\begin{figure}
	\begin{center}
		\includegraphics[width=80mm]{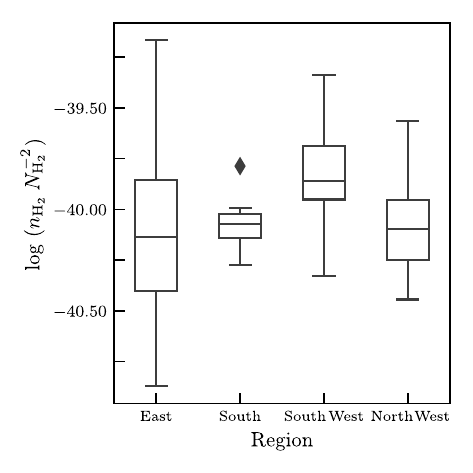}
	\end{center}
	\caption{Box plots for $n_\mathrm{H_2}N_\mathrm{H_2}^{-2}$, which is proportional to the virial parameter, of molecular clouds for each subregion. The diamond symbol indicates an outlier.}
	\label{fig:virial}
\end{figure}

Our results suggest no particular events triggering star formation externally in the southwestern and northwestern subregions, as neither $T_\mathrm{kin}$ nor $n_\mathrm{H_2}$ is enhanced there.
Moreover, it has been noted that the velocity dispersion of molecular gas is large in the western part of the ring \citep{Sato2021}.
These facts suggest that a virial parameter $\alpha_\mathrm{vir}$, which is the ratio of the kinematic energy to gravitational potential energy for clouds, at the southwestern subregion is larger than other subregions, which means star formation is suppressed there rather than triggered.
The virial parameter $\alpha_\mathrm{vir}$ can be evaluated using physical quantities in this paper with the following relation:
\begin{eqnarray}
	\alpha_\mathrm{vir} &\propto& \frac{\sigma^{2} R}{M} \nonumber \\
	&\propto& n_\mathrm{H_2}(\mathrm{d}N/\mathrm{d}v)^{-2} \nonumber \\
	&\propto& n_\mathrm{H_2}N_\mathrm{H_2}^{-2},
\end{eqnarray}
where $\sigma$ and $R$ are velocity dispersion and radius of a cloud, respectively.
Note that d$N$/d$v$ is a parameter directly obtained with the 3D analysis, which is converted to $N_\mathrm{H_2}$ (see subsection \ref{subsection:Method:HBI} and table \ref{tab:modelparam}).
Figure~\ref{fig:virial} shows a boxplot of $n_\mathrm{H_2}N_\mathrm{H_2}^{-2}$ for each voxel.
The values of $n_\mathrm{H_2}N_\mathrm{H_2}^{-2}$ in the southwestern subregion are predominantly larger than those in the other regions by 0.5 dex yet with a significant scatter.
Therefore, we cannot identify particular reasons to assume star-forming triggers in the southwestern subregion.  

The other possibility, i.e., deficiency of molecular gas, is more likely. 
While molecular gas is continuously supplied to the eastern subregion through the north-bar end, only insufficient gas supply is present in the southwestern subregion.
Even in diffuse molecular gas traced by $^{13}$CO integrated intensity, one can see a few 100~pc gap between the southern bar and the star-forming ring \citep[see, Figure 2 in][]{Sato2021}. 
This suggests that there is currently no significant gas supply from the southern bar to the star-forming ring.
Indeed, $\Sigma_{\rm H2}$ in the southwestern subregion is lower than in the eastern and other subregions.   
Therefore, we could reasonably assume that the high SFE in the southwestern region results from decreased gas mass rather than increased SFR.

\subsection{Popcorn Scenario vs. Pearls-on-a-string Scenario}
As described in Section \ref{sec:intro}, star formation in the star-forming rings can be explained with two different scenarios: ``popcorn'' and ``pearls-on-a-string'' scenarios.
In NGC\,613, sequential star formation is observed from the southern to southwestern subregions \citep{Falcon-Barroso2014}, which supports the ``pearls-on-a-string'' scenario. 
However, the eastern subregion has multiple star-forming peaks, which do not indicate age differences.
In addition, a star-forming peak is found in the northwestern subregion. 
Therefore, a simple pearls-on-a-string scenario cannot explain these observations; popcorn-like star formation is preferable for the eastern and northwestern subregions.
The large variation in the virial parameter in the eastern subregion (figure \ref{fig:virial}) suggests that the progression of star formation varies significantly from one molecular cloud to another.
This fact is consistent with the scenario that popcorn-like star formation is occurring in the eastern subregion.
That is, pearls-on-a-string-like and popcorn-like star formation modes appear to coexist in the NGC\,613 ring.

We may relate the difference in the SFE between the northwestern and eastern subregions to these two star-formation modes.
Assuming galactic rotation \citep{Miyamoto2017}, the time spent travelling from the southern to the southwestern subregion is expected to be a few Myr long, enough for star formation to continue.
A rich amount of gas is continuously supplied to the eastern region through the northern bar end.
Therefore, active star formation is sustained in a relatively narrow area of the eastern part of the ring, in which the age sequence between the individual star-forming regions cannot be identified.
In this case, star formation should appear popcorn-like.
Meanwhile, a further supply of gas is absent in downstream; hence, star formation continues consuming the remaining gas while traveling toward the northwestern region.  
In this case, we could identify the pearls-on-string-like age sequence of formed stars.
Thus, we may explain the star-forming properties in the NGC\,613 ring by assuming a hybrid of a stochastic (popcorn scenario) and a continual (pearls-on-a-string scenario) star formation.
Indeed, such a combination of the two star-formation scenarios in the star-forming ring was previously reported for NGC\,4736 \citep{vanderLaan2015}.
Our result suggests that star formation will fit a popcorn scenario if gas is continuously supplied.
If there is insufficient or intermittent gas supply, star formation will fit the pearls-on-a-string scenario.
The regions where the stochastic star formation occurs show higher SFE than the continual star-forming regions.

Such hybrid star formation in star-forming rings has been observed in recent simulations.
\citet{Moon2022} simulated asymmetric mass inflows into the ring. 
When gas continues to flow in from both sides of the bar, popcorn-like star formation occurs throughout the ring. 
However, if the inflow occurs rapidly on only one side, the inflowing gas flows to the other side of the ring, resulting in pearls-on-a-string star formation.
The gas then spreads out as it travels around the ring and returns to popcorn-type star formation. 
This explanation is also consistent with the coexistence of a stochastic and a continual star formation in NGC\,613.

\section{Summary}\label{sec:summary}

We presented 3D distributions of the volume density, column density, and kinetic temperature of molecular gas at a resolution of 0$\farcs$8 or $\sim$68~pc in the central $\sim$1.3 kpc region of the nearby active galaxy NGC\,613.
We used eight molecular lines ($^{13}$CO(1--0), C$^{18}$O(1--0), HCN(4--3), HCN(1--0), HCO$^{+}$(4--3), HCO$^{+}$(1--0), CS(7--6), and CS(2--1)) observed with ALMA. 
We exploited a non-LTE method with HB inference to derive physical quantities from these multiple-molecular-line observations.
This method successfully suppressed the artificial correlations between parameters created by systematic errors due to the limitations of the simple one-zone excitation analysis and calibration uncertainty. 
Our key findings are summarized as follows:

\begin{enumerate}
	
	\item The derived ranges of the volume densities $n_\mathrm{H_2}$ and kinetic temperatures $T_\mathrm{kin}$ are 10$^{3.21-3.85}$~cm$^{-3}$ and 10$^{2.33-2.64}$~K, respectively.
	We also obtained column densities $N_{\rm H_2}$ or the gas surface density $\Sigma_\mathrm{H_2}$ from the mean in the velocity direction, ranging from $N_\mathrm{H_2}$ = 10$^{20.8}$ to 10$^{22.1}$~cm$^{-2}$.  
	
	\item We examined the correlation between $\Sigma_\mathrm{H_2}$ and the star formation rate $\Sigma_\mathrm{SFR}$ obtained from the 110 GHz continuum flux map. 
	The inference shows that the molecular gas in the central region of NGC\,613 tends to exhibit an elevated $\Sigma_\mathrm{SFR}$ for the $\Sigma_\mathrm{H_2}$ value compared with both the galaxy-scale and 100 pc-scale Kennicutt--Schmidt law in nearby star-forming galaxies. 
	
	\item We investigated the physical origin of the departure from the known Kennicutt--Schmidt law in typical star-forming galaxies by dividing the data into four representative regions in the observed map based on the $^{13}$CO(1--0) integrated intensity image, i.e., east, south, southwest, and northwest. 
	We observed two distinct sequences between the $\Sigma_\mathrm{H_2}$-$\Sigma_\mathrm{SFR}$ diagram, where molecular clouds in the southwestern subregion exhibit the $\Sigma_\mathrm{SFR}$ value of $\sim$0.5~dex higher than that of the eastern subregion clouds. 
	
	\item We examined the origin of high SFEs in the southwestern clouds. 
	No systematic difference in volume densities, often considered a driver of the SFE variation, was observed between the clouds in the southwestern and other regions.
	We suggest that the amount of molecular gas in the southwestern region decreased, owing to rapid gas consumption resulting from star formation or a lack of gas supply.  
	
	\item This study was the first attempt to apply the non-LTE method with HB inference to external galaxies.
	The similar results in comparison with previous studies indicate that this method can be extended to other galaxies.
	
\end{enumerate}

\begin{acknowledgments}
	This study makes use of the following ALMA data: ADS/JAO.ALMA\#2013.1.01329.S and 2015.1.01487.S.
	ALMA is a partnership of ESO (representing its member states), NSF (USA), and NINS (Japan) together with NRC (Canada), NSC and ASIAA (Taiwan), and KASI (Republic of Korea) in cooperation with the Republic of Chile.
	The Joint ALMA Observatory is operated by ESO, AUI/NRAO, and NAOJ.
	Data analysis was partly carried out on the common-use data analysis computer system at the Astronomy Data Center (ADC) of the National Astronomical Observatory of Japan.
	This work was supported by JSPS KAKENHI Grant Number 20H00172 and the NAOJ ALMA Scientific Research Grant Number 2020-15A.
\end{acknowledgments}

\vspace{5mm}
\facility{ALMA}

\software{astropy \citep{astropy},  
	APLpy \citep{APLpy},
	CASA \citep{CASA2022}
}

\appendix
\restartappendixnumbering
\section{HCN(\texorpdfstring{$J$}{J} = 1--0)/\texorpdfstring{$^{13}$}{13}CO(\texorpdfstring{$J$}{J} = 1--0) line ratio along the horizontal axis}\label{sec:appendix_line_ratio}
\begin{figure}
	\begin{center}
		\includegraphics[width=80mm]{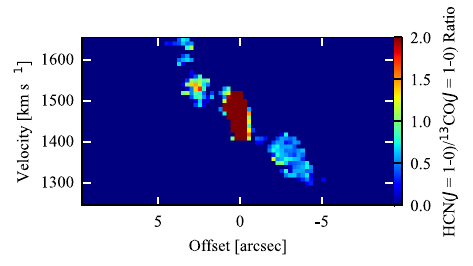}
	\end{center}
	\caption{The position-velocity diagram of HCN($J$ = 1--0)/$^{13}$CO($J$ = 1--0) line ratio along the horizontal axis. The positions where $^{13}$CO($J$ = 1--0) or HCN($J$ = 1--0) are below 3$\sigma$ are masked. The offset of 0$^{\prime\prime}$ corresponds to the galactic center of NGC\,613.} \label{fig:line_ratio}
\end{figure}

We produce an HCN($J$ = 1--0)/$^{13}$CO($J$ = 1--0) line ratio cube, and apply a mask based on the 3$\sigma$ threshold of $^{13}$CO($J$ = 1--0) (with 1$\sigma$ = 8.04\,$\times$\,10$^{-2}$ K) or HCN($J$ = 1--0) (with 1$\sigma$ = 2.00\,$\times$\,10$^{-1}$ K).
Next, we generate a PVD along the right ascension (horizontal axis), with the offset of 0 arcsec corresponding to the galactic center.
As depicted in figure~\ref{fig:line_ratio}, the PVD of HCN($J$ = 1--0)/$^{13}$CO($J$ = 1--0) line ratio displays a varying trend along the velocity direction (i.e., the vertical axis).
This outcome indicates that the physical properties of molecular clouds cannot be accurately estimated using the inference based on integrated intensity images.

\section{Derivation of Gas Properties using Hierarchical Bayesian Inference}\label{sec:appendix_HB}
The HB method calculates the hierarchical posterior probability \PDF{\pv,\Emv,\hpv|\Iobsv}, which is the simultaneous probability of the 3D distributions of the parameters (\pv) and nonstatistical errors of the input line intensities (\Emv), and hyperparameter (\hpv), under the condition that the distributions of the observed line intensities \Iobsv\ are known.  
The hyperparameter \hpv\ is introduced to determine the optimal shapes of the prior distributions of \pv\ and \Emv.
The posterior function is calculated from the likelihood function \PDF{\Iobsv|\pv,\Emv} using Bayes' theorem: 
\begin{eqnarray}
	\PDF{\pv,\Emv,\hpv|\Iobsv} & = & 
	\frac{
		\PDF{ \Iobsv|\pv,\Emv} \cdot \PDF{\pv,\Emv|\hpv} \cdot \PDF{\hpv}
	}
	{
		\PDF{\Iobsv} 
	}.
	\label{EQ3_0}
\end{eqnarray}
The second and third factors in the numerator are the prior and hyperprior distribution functions, respectively; their specific forms are determined based on statistical or physical modeling.

The likelihood function \PDF{\Iobsv|\pv,\Emv}\ was formulated with the assumption that the $j$th line intensity at the $i$th voxel follows a normal distribution with a mean of $I_{i,j}$ and a scale of $\delta_j$: 
\begin{eqnarray}
	\PDF{\Iobsv|\pv,\Emv} & = &
	\prod_{i,j}\frac{1}{\delta_{j}}\exp\left[- \frac{1}{2}\left(\frac{\Iobs_{i,j} - \Em_{i,j}\cdot\Icalci_{j}}{\delta_{j}}\right)^2\right], \label{EQ2}
\end{eqnarray}
where $\Em_{i,j}$ denotes the element of \Emv\ for the $j$th line at the $i$th voxel.  
The function $F(\pvi)_j$ provides the $i$th-line intensity of the model, calculated from the input parameter set at the $i$th voxel, \pvi. 
The rms noise levels of the ALMA maps were used for $\delta_j$.  

The prior distribution \PDF{\pv,\Emv|\hpv}\ was assumed to be the product of the log-normal function of \Emv\ and the multivariate student function of \pv, the function forms of which are given in equations 6 and 8 of T18, respectively.
The hyperparameter \hpv\ consists of the location ($\pv_0$) and scale matrix ($\Sigma$) of the prior \pv\ distribution and the scale parameters (\smv) of the prior \Emv\ distribution.
The same hyperprior modeling used in T18 was adopted in our analysis, except for the following two minor modifications:
\begin{description}
	\item[Addition of constraints on the prior correlation coefficients ]
	T18 used logistic hyperpriors to restrict the prior correlation coefficients ($R_{i,j} \equiv \Sigma_{i,j}/\sqrt{\Sigma_{i,i}\cdot\Sigma_{j,j}}$) to be nonnegative for the parameter pairs ($N_\mathrm{H_2}$, $n_\mathrm{H_2}$) and ($\phi$, $N_\mathrm{H_2}$), thereby forbidding artificial anticorrelation between them.  
	In addition, we also assumed the same logistic priors for ($N_\mathrm{H_2}$, $T_\mathrm{kin}$) and ($T_\mathrm{kin}$, $n_\mathrm{H_2}$) in the present analysis.  
	This was necessary because $T_\mathrm{kin}$, $N_\mathrm{H_2}$ and $n_\mathrm{H_2}$ degenerate faster in the likelihood function than in the T18 analysis because our input dataset includes fewer lines.
	We also limit all $R_{ij}$ elements to the range [$-0.8$, +0.8] to ensure the Markov Chain Monte Carlo (MCMC) covers a sufficiently wide parameter range within a reasonable computational time; if a correlation coefficient becomes $\sim$1, the prior probability \PDF{\pv,\epsilon|\theta} becomes highly sensitive to small changes in $\pv$, resulting in a small MCMC step size and hence a long computational time required for the convergence.
	\item[Addition of the lower limits to \smv ]
	T18 also applied logistic hyperpriors to \smv\ to set upper limits on its elements.  
	The present analysis used a modified hyperprior $\PDF{\smv} = \prod_j f_\mathrm{log}\left(\sm_j - \sigma_\mathrm{min}; a\right)\cdot f_\mathrm{log}\left(\sm_j - \sigma_\mathrm{max}; -a\right)$, where $f_\mathrm{log}(x;a)$ is a logistic function with a scale parameter $a$, that is, all elements of \smv\ were both lower- and upper-limited.  
	The values of $\sigma_\mathrm{min,max}$ were chosen to be 0.1 and 0.3, respectively.
\end{description}

The MCMC method was used to calculate the posterior distribution numerically.
The details of the sampling method are provided in T18.
The PPV cubes of $N_\mathrm{H_2}$, $n_\mathrm{H_2}$, $T_\mathrm{kin}$, and $\xmol\left(X\right)$ were created based on their marginal posterior probabilities, obtained by integrating \Emv, \hpv, and all nontarget \pv\ elements from the posterior distribution at every voxel.   
The median value of the marginal posterior was adopted as the voxel value when the 25th-75th percentile interval width was less than 0.2; otherwise, the voxel was left blank.  
Therefore, the relative uncertainty of the final result is better than 26\% for all voxels.
The number fraction of the filtered-out voxels is 1.2\%, indicating that the bias caused by the final filtering would be negligible.

We note a caveat on the hyperprior functions to enforce nonnegative $R_{ij}$ elements for the ($T_\mathrm{kin}$, $N_\mathrm{H_2}$) and ($T_\mathrm{kin}$, $n_\mathrm{H_2}$) parameter pairs.
Unlike the same constraint on the ($n_\mathrm{H_2}$, $N_\mathrm{H_2}$) pair, they are arbitrary assumptions introduced for a computational reason rather than based on ISM physics.
As mentioned in subsection \ref{subsection:Method:HBI}, they may hide true anticorrelation present in the real molecular clouds.
Observational results suggest, however, the absence of strong anticorrelation of $T_\mathrm{kin}$ with $N_\mathrm{H_2}$ and $n_\mathrm{H_2}$ in dense ISM.
We consider it reasonable to assume the same applies to the NGC\,613 ring.

\section{Comparison between hierarchical Bayesian and classical Bayesian models}\label{sec:appendix_comparison}
\begin{figure}
	\begin{center}
		\includegraphics[width=160mm]{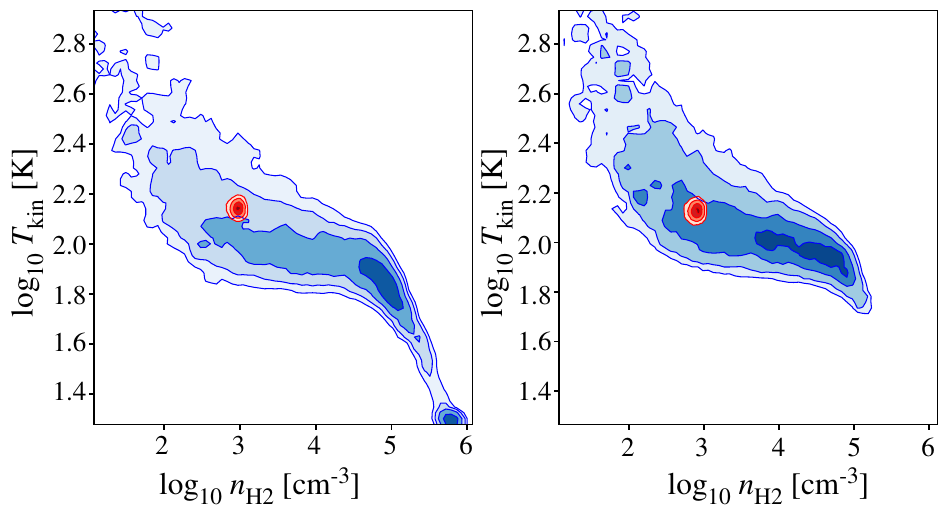}
	\end{center}
	\caption{Plots of volume density and gas kinematic temperature at two positions that show peaks of column density. Blue and red contours correspond to nonhierarchical and hierarchical Bayesian models, respectively. Contour levels illustrate the confidence intervals of 30\%, 50\%, 75\%, and 95\%.} \label{fig:HBvsNonHB}
\end{figure}

We calculate the molecular hydrogen volume density and gas kinematic temperature using the classical (i.e., nonhierarchical) Bayesian model to show the validity of the HB method.
The classical Bayesian model corresponds to the simple one-zone non-LTE model calculations with standard maximum-likelihood analysis.
The calculations were performed at two pixels that show peak column densities in the hierarchical Bayesian model.
Other parameters and assumptions are the same as in the HB method in the calculations.
Figure~\ref{fig:HBvsNonHB} represents plots of derived volume density and gas kinematic temperature.
The nonhierarchical Bayesian model cannot determine the optimal parameters, while the HB model infers a narrow range of hydrogen volume density and gas kinematic temperature in a narrow range. 
This result illustrates that the HB model is useful for calculating gas properties.

\bibliography{n613}{}
\bibliographystyle{aasjournal}

\end{document}

%% file: include.tex
\newcommand\myvector[1]{\ifmmode{\mbox{\boldmath ${#1}$}}\else{\boldmath {${#1}$}}\fi}
\newcommand{\Rt}{\ifmmode{{R_{13}}}\else${R_{13}}$\fi}
\newcommand{\Iobs}{\ifmmode{I}\else${I}$\fi}
\newcommand{\Icalc}{\ifmmode{{F}\left(\pv\right)}\else${{F}\left(\pv\right)}$\fi}
\newcommand{\Icalci}{\ifmmode{{F}\left(\pv_{i}\right)}\else${{F}\left(\pv_{i}\right)}$\fi}
\newcommand{\Icalcv}{\ifmmode{\myvector{F}\left(\pv\right)}\else${\myvector{F}\left(\pv\right)}$\fi}
\newcommand{\xmol}{\ifmmode{{x_{\rm mol}}}\else${x_{\rm mol}}$\fi}
\newcommand{\xmolp}[1]{\ifmmode{{x_{\rm mol}\left(\mathrm{#1}\right)}}\else${x_{\rm mol}\left(\mbox{#1}\right)}$\fi}
\newcommand{\ff}{\ifmmode{\Phi}\else${\Phi}$\fi}
\newcommand{\fff}{\ifmmode{\phi}\else${\phi}$\fi}
\newcommand{\fcal}{\ifmmode{f_{\rm cal}}\else${f_{\rm cal}}$\fi}
\newcommand{\Iobsv}{\myvector{\Iobs}}

\newcommand\unit[1]{\ifmmode{\mbox{\boldmath $e$}_{#1}}\else{${\boldmath e}_{#1}$}\fi}
\newcommand{\Ea}{\ifmmode\epsilon^{\rm a}\else$\epsilon_{\rm a}$\fi}
\newcommand{\Em}{\ifmmode\epsilon\else$\epsilon$\fi}

\newcommand{\Emv}{\myvector{\Em}}
\newcommand{\pv}{\myvector{p}}

\newcommand{\pvi}{\myvector{p_i}}

\newcommand{\PDF}[1]{{\ifmmode \mathrm{Pr}\left(#1\right) \else $\mathrm{Pr}\left(#1\right)$ \fi}}
\newcommand{\hpv}{\myvector{\theta}}
\newcommand{\sa}{\ifmmode\sigma\else$\sigma$\fi}
\newcommand{\sm}{\ifmmode\sigma\else$\sigma$\fi}
\newcommand{\scal}{\ifmmode\sigma_{\rm cal}\else$\sigma_{\rm cal}$\fi}
\newcommand{\smv}{\myvector{\sm}}